\documentclass[iop]{emulateapj}
\pdfoutput=1

\usepackage{natbib}
\usepackage{threeparttable}
\usepackage{amsmath}
\usepackage{graphicx}
\usepackage{subfigure}

\newcommand{\kms}{\mbox{km~s$^{-1}$}}

\newcommand{\cm}{\mbox{cm$^{-2}$}}

\newcommand{\av}{\ensuremath{\mbox{$A_{\rm V}$}}}

\newcommand{\jnu}{\ensuremath{\mbox{$J_\nu(z)$}}}
\newcommand{\nnu}{\ensuremath{\mbox{$N_\nu(z)$}}}
\newcommand{\tdust}{\ensuremath{\mbox{$T_{\rm dust}$}}}
\newcommand{\yc}{\ensuremath{\mbox{$y_{\rm C}$}}}
\newcommand{\yd}{\ensuremath{\mbox{$y_{\rm d}$}}}

\newcommand{\tcmb}{\ensuremath{\mbox{$T_{\rm CMB}$}}}
\newcommand{\ndust}{\ensuremath{\mbox{$N_{\nu,{\rm dust}}$}}}
\newcommand{\idust}{\ensuremath{\mbox{$I_{\nu,{\rm dust}}$}}}
\newcommand{\zmax}{\ensuremath{\mbox{$z_{\rm max}$}}}

\shortauthors{Imara \& Loeb}

\begin{document}

\title{The Distortion of the Cosmic Microwave Background Spectrum Due to Intergalactic Dust}

\author{Nia Imara \& Abraham Loeb}
\affil{Harvard-Smithsonian Center for Astrophysics, 60 Garden Street, Cambridge, MA 02138}

\email{nimara@cfa.harvard.edu}

\begin{abstract}
Infrared emission from intergalactic dust might compromise the ability of future experiments to detect subtle spectral distortions in the Cosmic Microwave Background (CMB) from the early Universe.  We provide the first estimate of foreground contamination of the CMB signal due to diffuse dust emission in the intergalactic medium.
We use models of the extragalactic background light to calculate the intensity of intergalactic dust emission and find that emission by intergalactic dust at redshifts $z\lesssim 0.5$ exceeds the sensitivity of the planned Primordial Inflation Explorer (PIXIE) to CMB spectral distortions by 1--3 orders of magnitude.  We place an upper limit of $0.23\%$ on the contribution to the far-infrared background  from intergalactic dust emission.  
\end{abstract}



\section{Introduction}
The standard model of cosmology predicts small distortions in the Planckian spectrum  of the cosmic microwave background (CMB) due to processes that heat, cool, scatter, and generate CMB photons over the history of the Universe; Sunyaev \& Zeldovich 1970; Zeldovich \& Sunyaev 1969; Sunyaev \& Zeldovich 1970; Illarionov \& Sunyaev 1975;  Hu \& Silk 1993; Burigana \& Salvaterra 2003; Sunyaev \& Chluba 2009; Chluba \& Sunyaev 2012; Sunyaev \& Khatri 2013).   Measurements by the FIRAS (Far InfRared Absolute Spectrometer) instrument on-board the Cosmic Background Explorer (COBE) satellite demonstrated that the CMB spectrum is a perfect blackbody with a temperature $\tcmb=2.725\pm 0.001$ K and limited deviations from a blackbody spectrum to $\Delta I_\nu \lesssim 10^{-5}$  (Mather et al. 1994; Fixsen et al. 1996; Fixsen 2009).  Since then, a main goal in astrophysical cosmology has been to determine precisely how much the CMB departs from a perfect blackbody spectrum.  Observations of such spectral distortions would help constrain inflationary models and yield other important information about the early history of the Universe, including recombination of hydrogen and helium at redshifts $z\sim 1100$--6000, the formation of the first stars, and the epoch of reionization (e.g., Chluba \& Sunyaev 2012; Chluba et al. 2012; Chluba 2013, 2016; Chluba \& Jeong 2014; Sunyaev \& Khatri 2013).  

At redshifts $z\gtrsim 2\times 10^6$, thermalization processes erased distortions of the CMB.   The blackbody spectrum was sustained by processes including Compton scattering, Bremsstrahlung (Zeldovich \& Sunyaev 1970; Illarionov \& Sunyaev 1975a; Illarionov \& Sunyaev 1975b), and double Compton scattering (Danese \& de Zotti 1982; Burigana et al. 1991).  As the Universe expanded and cooled, thermalization processes gradually became less efficient.  By $z\lesssim 10^6$, energy released into the Universe was capable of generating deviations, or distortions, into the CMB spectrum that may be observed today.  

The epoch of energy release dictates the kind of spectral distortions induced in the CMB spectrum.  They are generally categorized into two types, named $\mu$- and $y$-type distortions.  The former represents a frequency-dependent chemical potential, $\mu(\nu)$, that develops around $z\gtrsim 10^5$, when Compton processes bring photons into complete kinetic equilibrium with electrons.  Under these circumstances, the CMB spectrum is described by a Bose-Einstein distribution, and any energy release yields a chemical potential described by $\mu(\nu)$ (e.g., Chluba \& Sunyaev 2012).  At lower redshifts ($z\lesssim 3\times 10^3 - 10^4$), $y$-type distortions, also called Compton-$y$ distortions, can form since the efficiency of Compton processes decreases, and so  kinetic equilibrium between photons and electrons is no longer realized.  Thus, due to incomplete thermalization of photons and electrons, low-frequency photons undergo small amounts of up-scattering, producing a $y$-type distortion. 

To date, the spectral distortions have not been detected, although COBE/FIRAS has constrained them to $|y|\lesssim 10^{-5}$ and $|\mu|\lesssim 10^{-4}$ (Mather et al. 1994; Fixsen et al. 1996).  Recently, however, new missions called the Primordial Inflation Explorer (PIXIE; Kogut et al. 2011) and the Polarized Radiation Imaging and Spectroscopy Mission (PRISM; Andr\'e et al. 2014) have been proposed that would exceed the spectral sensitivity limits of COBE/FIRAS by 3--4 orders of magnitude.  The greatly improved sensitivity of PIXIE will enable detections of CMB spectral distortions with $y\sim 10^{-8}$ and $\mu \sim 5\times 10^{-8}$ at the $5\sigma$ level (Kogut et al. 2011). PRISM is expected to detect $\mu$- and $y$-type distortions of few$\times 10^{-9}$ (Andr\'e et al. 2014; Tashiro 2014).

\begin{figure}[t!]
\epsscale{1.2}
\plotone{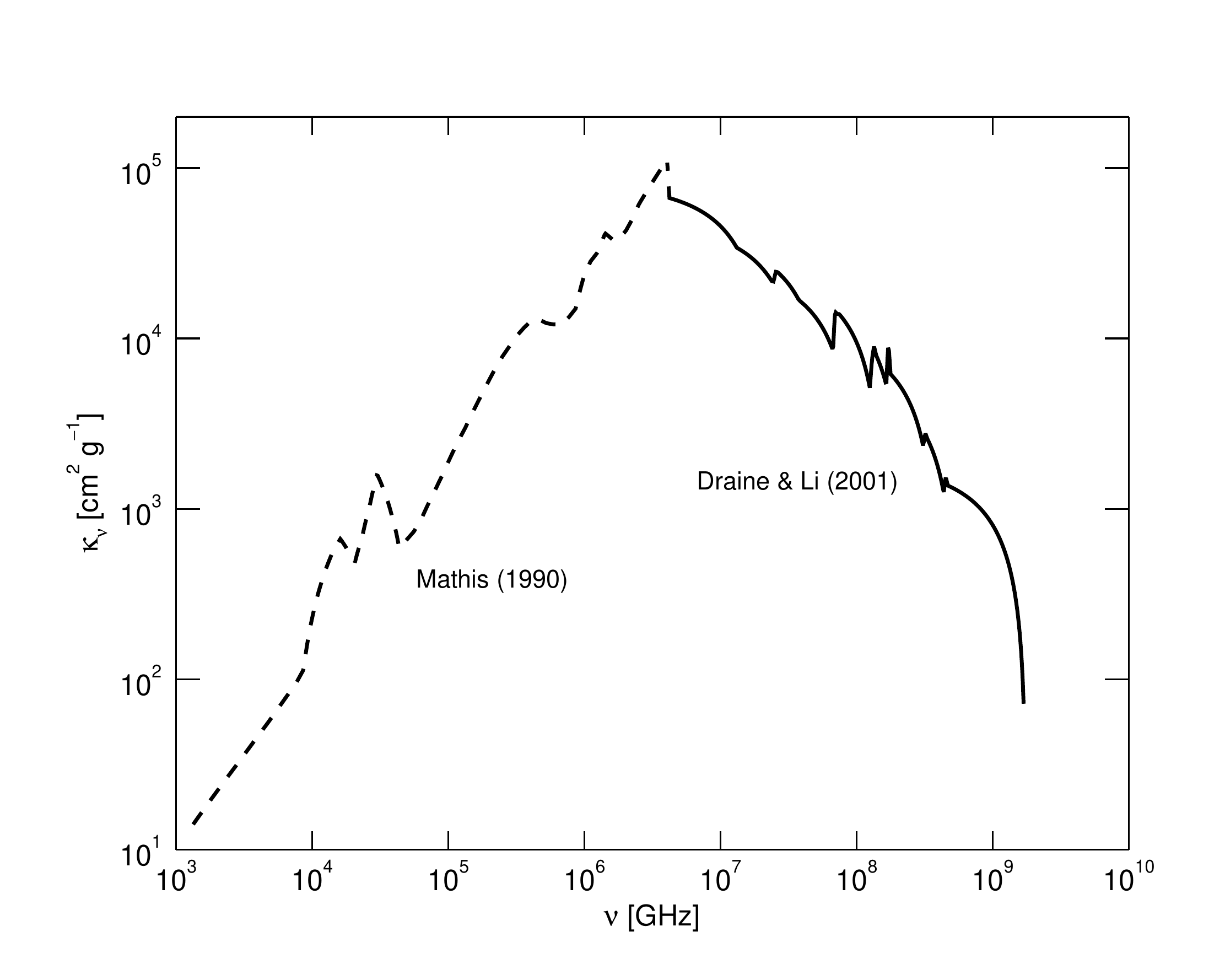} 
\caption{Dust opacity, $\kappa_\nu$, versus frequency, $\nu$, based on the mean extinction law of the Galaxy of Mathis (1990; dashed line) and Draine \& Li (2001; solid line).\label{fig:kappa}}
\end{figure}

In anticipation of these missions, it is important to consider all possible sources of foreground contamination that could mask the signals we hope to detect.  For instance, the PIXIE collaboration has focused on characterizing the dominant foregrounds of the Galactic interstellar medium (ISM), in particular, polarized synchrotron radiation and thermal dust emission (Kogut et al. 2011).  They determined that the Galactic foregrounds and CMB emission could be disentangled due to their different frequency spectra (see Fig. 4 of Kogut et al. 2011) and distribution on the sky.   

Extragalactic foreground contamination from continuum and spectral line emission is also a cause for concern.  For instance, the diffuse CO background from star-forming galaxies occupies the same frequency-space as does CMB emission.  Righi, Hern\'{a}ndez-Monteagudo, \& Sunyaev (2008) and de Zotti et al. (2015) estimated the signal of the redshift-integrated CO emission.  Under different assumptions, both groups found that the CO background signal is considerably higher than the PIXIE sensitivity, over various frequency ranges.  Recently, Mashian et al. (2016) used large-velocity gradient (LVG) modeling to produce the complete CO spectral line energy distribution from galaxies throughout cosmic history.  They demonstrated that the cumulative CO spectrum from star-forming galaxies, between $z\sim 15$ and the present, lies 1--3 orders of magnitude above the PIXIE sensitivity to $\mu$- and $y$-type distortions in the frequency range 30--300 GHz.

To date, little attention has been given to the contamination of CMB distortions due to emission by intergalactic dust, which is the focus of this work.  Analogous to the reddening and attenuation of starlight due to interstellar dust in galaxies, light spanning ultraviolet (UV) to infrared (IR) frequencies is absorbed and scattered by dust grains as it passes through a dust-enriched intergalactic medium (IGM).  Outflows from stars, supernova explosions, and active galactic nuclei winds are all mechanisms that may contribute to the expulsion of dust from galaxies into the IGM.  The enrichment of the IGM by heavy elements has been modeled by a number of studies (e.g., Cen \& Ostriker 1999; Aguirre et al. 2001; Theuns et al. 2002; Furlanetto \& Loeb 2003; Dav\'e \& Oppenheimer 2007; Dav\'e et al. 2011; Keating et al. 2014; Pallottini et al. 2014; Maio \& Tescari 2015).  Presumably, the same outflows that expel metals from galaxies also carry dust.  Observational evidence for intergalactic dust comes from dust extinction and metal line absorption measurements along the line of sight to background sources, including galaxies and quasars (e.g., Meurer et al. 1999; M\"ortsell \& Goobar 2003; Schaye et al. 2003; Gallerani et al. 2010;  M\'enard et al. 2010; Johansson \& M\"ortsell 2012; Bouwens et al. 2015).  

In this paper, we show that emission from intergalactic dust generates a substantial excess foreground that must be taken into account in order to reveal underlying CMB spectral distortions with the next generation of detectors.  In \S 2, we describe our method for calculating the IGM dust temperature and spectrum.  We present and discuss our results \S 3.  Section 4 provides concluding remarks and a summary of our main findings. Throughout, we assume standard values for the cosmological parameters: $H_0 = 70~\kms~\rm{Mpc}^{-1}$, $\Omega_b = 0.045$, $\Omega_m = 0.3$, and $\Omega_\Lambda =0.7$ (Planck Collaboration XI 2015).

\section{Model Description}
To begin, we want to quantify the number density of photons capable of being absorbed by dust grains in the IGM.  Following Wright (1981) and Loeb \& Haiman (1997),  we define the comoving number density of photons with a comoving frequency, $\nu$, as
\begin{equation}\label{eq:nnu}
N_\nu(z)=\frac{4\pi}{hc(1+z)^3} J_{\nu(1+z)}(z),
\end{equation}
where \jnu~is the specific intensity of the background radiation field (including microwave background, starlight, and dust emission) in erg $\cm~{\rm s}^{-1}~{\rm Hz}^{-1}~{\rm sr}^{-1}$. 

To describe the evolution of dust temperature, \tdust, with redshift, we assume that the dust is in thermal equilibrium with the total radiation field.  Under this premise, the power absorbed by dust in the UV is equal to the power it emits in the infrared, yielding the implicit equation,
\begin{equation}\label{eq:tdust}
\begin{split}
\int_0^\infty \left \{ \nnu - \frac{8\pi}{c^3} \frac{\nu^3}{{\rm exp}[h\nu(1+z)/k_{\rm B}\tdust]-1} \right\}\kappa_{\nu(1+z)}d{\nu}\\
- \int_0^\infty  \left \{\frac{8\pi}{c^3} \frac{\nu^3}{{\rm exp}[h\nu(1+z)/k_{\rm B} \tcmb]-1}   \right\} \kappa_{\nu(1+z)}d{\nu}=0.
\end{split}
\end{equation}

\begin{figure}[t!]
\epsscale{1.2}
\plotone{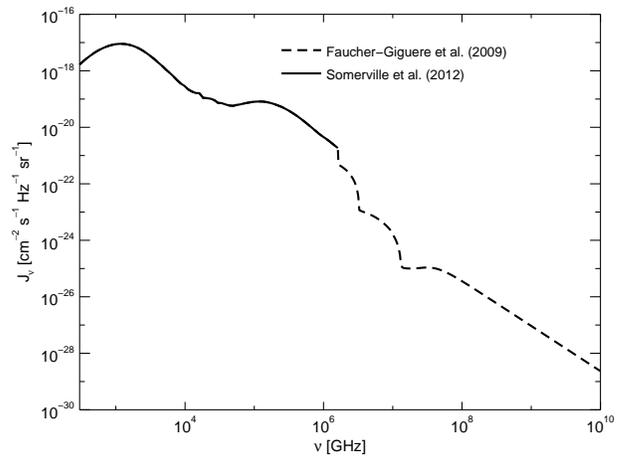} 
\caption{Adopted models of the specific intensity of the extragalactic background radiation, $J_\nu$, from Faucher-Gigu\`ere et al. (2009) and Somerville et al. (2009).
\label{fig:jnu}}
\end{figure}

We assume a single blackbody temperature for the dust, \tdust, excluding a mixture of dust grain temperatures.  Since the dust temperature at higher redshifts was warmer on average than the dust temperature today, $\tdust_{,0}$, we define $\tdust\equiv \tdust_{,0}(1+z)$.  We also consider the contribution of the CMB at a temperature, $\tcmb=2.725(1+z)$.  For the dust opacity, $\kappa_\nu$, in equation (\ref{eq:tdust}), we adopt the Galactic extinction laws of Mathis (1990) and Li \& Draine (2001).  We calculate an interpolated function for $\kappa_\nu$ based on a combination of these two models, shown in Figure 1, since the Mathis (1990) model extends to lower frequencies, down to $\sim 10^{12}$ Hz, while the Draine \& Li (2001) model extends up to $\sim 10^{18}$ Hz.  For frequencies below $\nu \le 10^{12}$ Hz, we follow Beckwith et al. (1990) and adopt a power law treatment of the opacity, 
\begin{equation}
\kappa_\nu=0.1 \left( \frac{\nu}{1000~\rm{GHz}} \right)^2~\rm{cm}^2 \rm{g}^{-1}.
\end{equation}

\subsection{Ionizing Background}
To determine \jnu~in equation (\ref{eq:nnu}), we adopt the model of Faucher-Gigu\`{e}re et al. (2009) for the ionizing background spectra at $z=0$--10 and the model of Somerville et al. (2012) for the extragalactic background light (EBL) at $z=0$.  Faucher-Gigu\`{e}re et al. (2009) calculated \jnu, for $\nu\gtrsim 1.6\times 10^{15}~\rm{Hz}$, at various redshifts by solving the cosmological radiative transfer equation.  For  $\nu\lesssim 10^{15}~\rm{Hz}$, where the EBL is dominated by emission from star-forming galaxies, we use the Somerville et al. (2012) model.  Since we do not know the exact shape of the EBL spectrum at $z>0$, we make the simplifying assumption that as $z$ increases, the spectrum only shifts in amplitude following the Faucher-Gigu\`{e}re et al. (2009) model. 

\begin{figure}
\epsscale{1.2}
\plotone{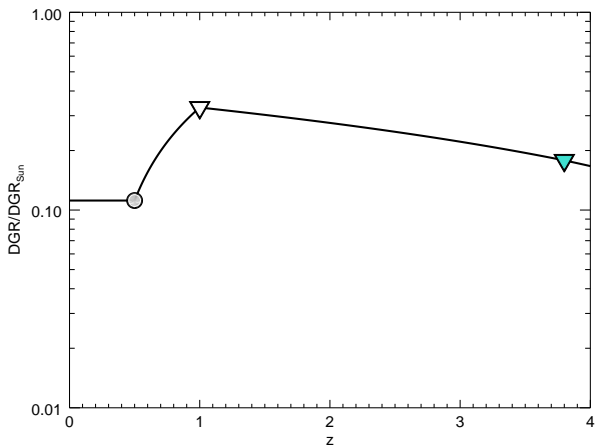} 
\caption{The intergalactic dust-to-gas ratio (DGR), with respect to $\rm{DGR}_{\rm Sun}=1/162$ (Zubko et al. (2004), versus redshift, $z$.  The filled and open triangles are upper limits derived from the observations of Bouwens et al. (2015) and Johansson \& Mortsell (2012), respectively.  The circle represents a measurement of the DGR derived from observations of M\'enard et al. (2010).  The curve through the data is an interpolation of DGR$(z)$. 
\vspace{3 mm}
\label{fig:dgr}}
\end{figure}
\newpage
\subsection{Intergalactic Dust Emission}\label{sec:yc}
The $y$-type distortion is proportional to the total energy injected into the CMB, $y=\frac{1}{4}\Delta E/E_\gamma$, where $E_\gamma=h\nu$ is the CMB energy density (e.g., Sunyaev \& Khatri 2013).  We are interested in an analogous quantity that only considers the foreground contribution from intergalactic dust emission.  We define an ``effective $y$-parameter'' \yd, comparing the number density of photons emitted by intergalactic dust to the number density of CMB photons:
\begin{equation}\label{eq:yc}
\yd\equiv \frac{1}{4} \left( \frac{\int d\nu N_{\nu,d} }{\int d\nu N_{\nu,0}} \right) , 
\end{equation}
where the integrals are evaluated over 30 GHz to 6 THz, the frequency range of PIXIE (Kogut et al. 2011) and PRISM (Andr\'e et al. 2014), and the number density of CMB photons at $z=0$ is
\begin{equation}
N_{\nu,0}=\frac{8\pi\nu^3/c^3}{{\rm exp}[h\nu(1+z)/k_{\rm B} \tcmb ]-1 }
\end{equation}
In equation (\ref{eq:yc}), $\ndust(z)$ is the comoving number density of photons due to dust emitting at temperature, \tdust.  To determine $\ndust(z)$, we employ the cosmological radiative transfer equation,
\begin{equation}\label{eq:Nnu}
\begin{split}
\frac{d\ndust (z_{\rm max})}{dz}&=\frac{c dt}{dz}\left[ \frac{4\pi}{hc}j_\nu(z) -\alpha_{\nu(1+z)} N_{\nu,0}   \right] \\
\ndust (z_{\rm max})  &=\int_0^{z_{\rm max}} \frac{c}{H(1+z)}  \\
   & \times \left[ \frac{4\pi}{hc} j_\nu(z) -\alpha_{\nu(1+z)} N_{\nu,0}   \right] dz  ,
\end{split} 
\end{equation}
since $(c dt/dz)=-c/[H(1+z)]$, with the Hubble parameter $H=H_0[\Omega_m (1+z)^3 +\Omega_\Lambda]^{1/2}$, and where $j_\nu(z)$ is the emission coefficient due to dust:
\begin{equation}\label{eq:jnu}
j_\nu(z)=\frac{2h\nu^3/c^2}{ {\rm exp}[h\nu(1+z)/k_{\rm B}\tdust]-1 } \alpha_{\nu(1+z)}.
\end{equation}
In the above expressions, the dust absorption coefficient, $\alpha_{\nu(1+z)}=\rho_{\rm dust}(z)\kappa_{\nu(1+z)}$, depends on the dust opacity, $\kappa_{\nu(1+z)}$, and the dust mass density, $\rho_{\rm dust}(z)$.  The latter is the product of the gas mass density,  $\rho_{\rm gas}(z)$, and a dust-to-gas ratio, DGR, as
\begin{equation}\label{eq:rho}
\begin{split}
\rho_{\rm dust}(z) &= \rho_{\rm gas}(z) \times {\rm DGR}(z)  \\
 &= \Omega_b\rho_{\rm crit} (1+z)^3 \times {\rm DGR}(z),
\end{split}
\end{equation}
where $\rho_{\rm crit}= 3H^2/(8\pi G)$ is the critical density.  The intensity of emission from intergalactic dust, \idust, is obtained from equation (\ref{eq:Nnu}) via $\idust=(2h\nu^3/c^2)\ndust$.

Knowledge of the evolution of the DGR is limited, since it requires measurements of the dust mass density at different epochs.  Such measurements are especially difficult to obtain at high redshifts due to challenges associated with disentangling the contributions to dust emission (or extinction) from various sources along a given line of sight. In Imara \& Loeb (2016), we constrained the DGR in the IGM at discrete redshifts by comparing our theoretical predictions of the optical depth due to dust with observations of dust extinction from the literature.  We estimated a DGR of $\sim 0.11$ at $z=0.5$, using optical extinction measurements of M\'enard et al. (2010) at the same redshift.  At higher redshifts, $1\lesssim z \lesssim 10$, we utilized upper limits of dust extinction observations to set upper bounds on the DGR.   Figure \ref{fig:dgr} displays a plot of DGR($z$) from $z=0$ to 4, obtained by interpolating through the discrete calculations of the DGR predicted in Imara \& Loeb (2016).  We use these interpolated values to solve for $\rho_{\rm dust}(z)$ in equation (\ref{eq:rho}).  We emphasize that for $z>0.5$, the curve in Figure \ref{fig:dgr} should be regarded as an upper limit to the DGR.  Therefore,  estimates of $\rho_{\rm dust}(z)$ for $z>0.5$ will be upper limits.  For $z<0.5$, we assume a constant DGR of 0.11, though we expect, in reality, the DGR in the IGM will increase approaching toward $z=0$.  Thus, calculations of $\rho_{\rm dust}(z<0.5)$ should be considered lower limits.

\begin{figure}[t!]
\epsscale{1.2}
\plotone{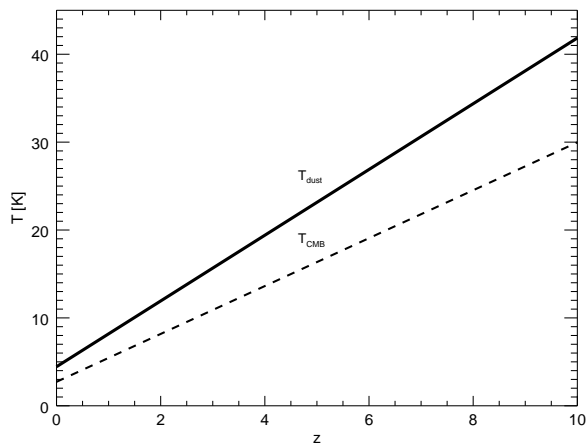} 
\caption{Redshift evolution of dust temperature, \tdust (solid line), and CMB temperature, \tcmb~(dashed line).
\vspace{2 mm}\label{fig:tdust}}
\end{figure}

\section{Results and Discussion}\label{sec:results}
\begin{figure}
\epsscale{1.2}
\plotone{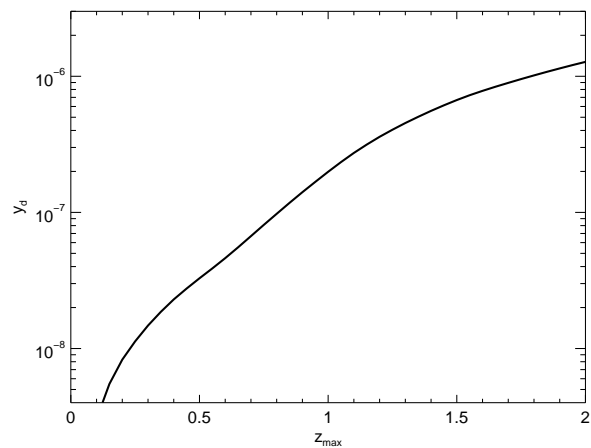} 
\caption{Redshift evolution of the Compton-$y$ parameter, \yc.\label{fig:ycz}}
\end{figure}

With models for the extragalactic background and the dust opacity, we solve equation (\ref{eq:tdust}) for \tdust~and plot the result in Figure \ref{fig:tdust}.  The plot also shows the history of the CMB temperature (dashed line).  Figure \ref{fig:ycz} displays the ``effective $y$-parameter,'' \yd, as a function of redshift, the result of numerically integrating Equations (\ref{eq:yc})--(\ref{eq:rho}).  In the following, we discuss the assumptions that influence our results (\S\ref{sec:unc}), and we forecast how the foreground emission due to intergalactic dust compares with the CMB spectral distortions that PIXIE and other future experiments hope to detect (\S\ref{sec:distort}).

\subsection{Assumptions and Uncertainties}\label{sec:unc}
Fortunately, our calculations should not be biased by the clustering of sources, since thermal dust emission is spectrally smooth.  Unlike estimates of the CO foreground, for instance, which has frequency fluctuations resulting from the clustering of sources on the sky (Mashian et al. 2016), the calculations performed here are not hampered by lack of knowledge of the spatial distribution of dust-emitting sources.

\begin{figure}
\epsscale{1.2}
\plotone{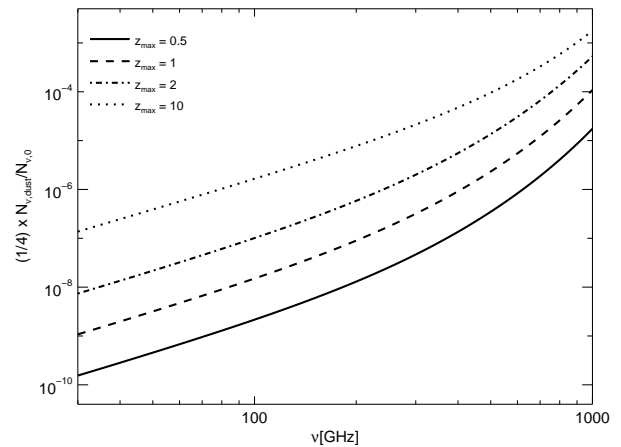} 
\caption{The excess of flux from intergalactic dust over the flux from CMB photons, $\frac{1}{4}\ndust/N_{\nu,0}$, as a function of frequency. The solid line represents a lower limit to the excess, only considering the contribution of dust from redshifts $z<0.5$.  The dashed, dash-dot, and dotted curves are upper limits obtained for dust at $z<1$, $z<2$, and $z<10$, respectively.  \label{fig:nu}}
\end{figure}

Our calculation of \yc, however, is contingent on a number of uncertain factors involving the intrinsic properties of intergalactic dust, such as its opacity and DGR.  For the dust extinction law, we assume a Galactic model consisting of a combination of silicate and carbonaceous grains, with a selective extinction of $R_{\rm V}\equiv \av/(A_{\rm B}-\av)=3.1$ (see Li \& Draine 2001). The selective extinction  defines the slope of the extinction curve in the optical band  from $0.125\lesssim \lambda\lesssim 3.5~\micron$.   Smaller grains tend to scatter light at shorter wavelengths, corresponding to a steeper slope and smaller values of $R_{\rm V}$.  Aguirre (1999) and Bianchi \& Ferrara (2005) propose that small dust grains are preferentially destroyed during the processes that expel dust from galaxies. If correct, and if intergalactic dust has a size distribution skewed towards larger grains than the Milky Way, this would suggest higher values of $R_{\rm V}$ than in the Milky Way.  On the other hand, if the rarefied, low-metallicity IGM is better described by dust with a grain size distribution more akin to a low-metallicity galaxy such as the Small Magellanic Cloud (SMC), this would correspond to $R_{\rm V}<3.1$. It has been suggested that the low-metallicity of the SMC contributes to the suppression of dust grain growth in the bar of this galaxy, in which $R_{\rm V}\approx 2.7$ (e.g., Bouchet et al. 1985; Gordon et al. 2003).

The uncertainties in the opacity, $\kappa_\nu$, and the DGR are directly proportional to changes in \yd~and are multiplicative.  That is, changing  $\kappa_\nu$ by a factor $f_\kappa$ or the DGR by a factor $f_{\rm DGR}$ results in a change to \yd~by a factor of  $f_\kappa f_{\rm DGR}$.  Nevertheless, in Imara \& Loeb (2016), we demonstrated that the estimated optical depth and DGR of the low-$z$ IGM are fairly robust to changes in the selective extinction.  We calculated the optical depth due to IGM dust averaged along arbitrary lines of sight and showed that at low redshifts, the derived optical depths---and thus, the derived DGRs---are nearly identical for  $3.1 \lesssim R_{\rm V}\lesssim 5.5$.

\begin{figure}
\epsscale{1.2}
\plotone{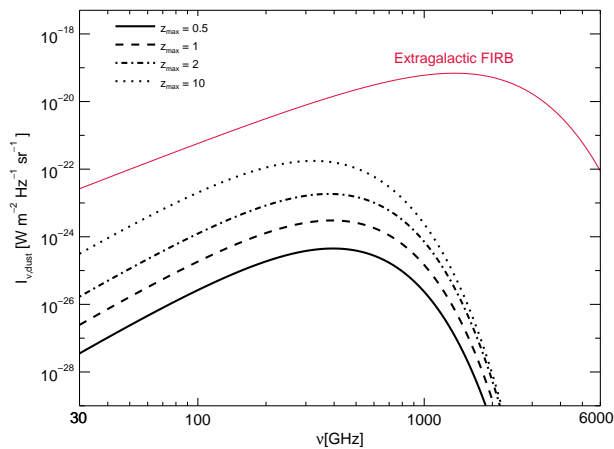} 
\caption{Black curves: Spectrum of intergalactic dust emission, \idust, as a function of observed frequency.  The solid black curve indicates a lower limit to \idust, obtained by considering dust emission from $z<0.5$.  The dashed and dash-dot curves are upper limits obtained for dust at $z<1$ and $z<2$, respectively.  The red curve delineates the extragalactic FIRB spectrum derived by Fixsen et al. (1998). \label{fig:inua}}
\end{figure}

\subsection{Detecting Spectral Distortions}\label{sec:distort}
In Figures \ref{fig:nu} and \ref{fig:inua} we present predictions of the foreground intergalactic dust emission between $z=0$ and 10.  In Figure \ref{fig:nu}, the amount of dust emission as a function of frequency is given in terms of $\ndust/N_{\nu,0}$, the ratio of number density of photons produced by dust emission to that of the CMB.  The dust emission is expressed in terms of intensity, \idust, in Figure \ref{fig:inua}.  The solid black curves in both figures represent lower limits obtained by considering the contribution of dust only up to redshift $\zmax = 0.5$.  These are lower limits for two main reasons.  First, our derivation of the dust emission spectrum for $z<0.5$ does not include the contribution from IGM dust at higher redshifts, for which we have only upper limits on the DGR.  Second, we assume a constant DGR between $z=0$ and 0.5, based on the M\'enard et al. (2010) detection of intergalactic extinction at $z=0.5$  (see \S\ref{sec:yc}).  If the DGR rises towards $z=0$, as might be expected since galaxies will have had more time to expel dust into the IGM, the foreground due to dust emission would also increase above the spectrum derived here.   In Figures \ref{fig:nu} and \ref{fig:inua}, we also show upper limits to the level of dust emission, derived for contributions of dust from $\zmax \le 1$ (dashed line), $\zmax \le 2$ (dash-dot line), and $\zmax \le 10$ (dotted line).

\begin{figure}
\epsscale{1.2}
\plotone{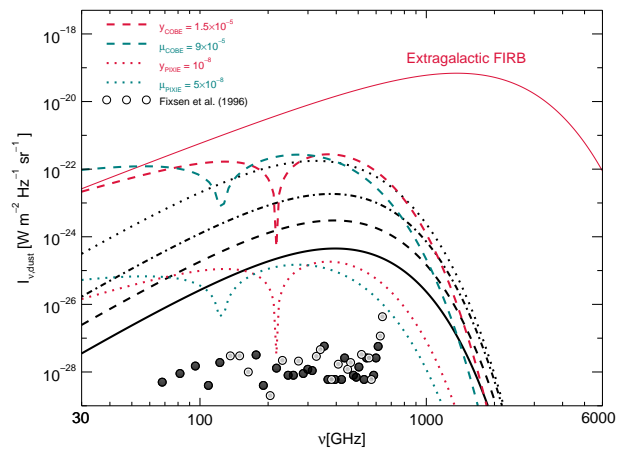} 
\caption{Black curves and solid red curve: same as Figure \ref{fig:inua}.  Overplotted are the absolute values of $\Delta I_\nu^\mu$ and $\Delta I_\nu^y$, given by equations (\ref{eq:inu_mu}) and (\ref{eq:inu_y}): corresponding to the COBE/FIRAS upper limits ($2\sigma$; dashed curves) and to the PIXIE sensitivity limits at the $5\sigma$ level (dotted curves).  The cusps in the $\Delta I_\nu^\mu$ curves mark the transition from negative to positive $\mu$-type distortions.  The circles denote the FIRAS-measured CMB residuals from Fixsen et al. (1996).  The filled circles have positive values; the open circles are the absolute value of negative residuals. \label{fig:inub}\vspace{2 mm}}
\end{figure}

In Figure \ref{fig:inub}, together with \idust, we show the deviations from the CMB blackbody spectrum as determined by Fixsen et al. (1996) from COBE/FIRAS measurements.  Also overplotted are the upper limit estimates on the $\mu$- and $y$-type distortions from COBE/FIRAS (dashed curves), and the PIXIE sensitivity to $\mu$- and $y$-type distortions at the $5\sigma$ level (dotted curves).   From the Fokker-Planck approximation of the Boltzmann equation with Compton scattering (the Kompaneets equation; Kompaneets 1957), the spectral distortions as a function of frequency may be written as (Zeldovich \& Sunyaev 1969; Sunyaev \& Khatri 2013; Chluba 2014):
\begin{equation}\label{eq:inu_mu}
\Delta I_\nu^\mu = \frac{2h\nu^3}{c^2}\times \mu \frac{e^x}{(e^x - 1)^2} \left(\frac{x}{2.19}-1\right)
\end{equation}
and
\begin{equation}\label{eq:inu_y}
\Delta I_\nu^y = \frac{2h\nu^3}{c^2}\times y \frac{xe^x}{(e^x - 1)^2}\left[ x \left(\frac{e^x+1}{e^x-1}\right)-4 \right]
\end{equation}
where $x=h\nu/(k_{\rm B}\tcmb)$ is the dimensionless frequency, $h$ is Planck's constant, and $k_{\rm B}$ is Boltzmann's constant.  

The instrument sensitivity of PIXIE is $\Delta I_\nu = 5\times 10^{-26}~\rm{Wm}^{-2}\rm{Hz}^{-1}\rm{Sr}^{-1}$ in each of the 400 frequency bins (Kogut et al. 2011).  This will allow PIXIE to detect distortions with $\mu=5\times 10^{-8}$ and $y=10^{-8}$  at the $5\sigma$ level (Kogut et al. 2011a; Chluba 2013; Chluba \& Jeong 2014). (Note that the anticipated sensitivity of PRISM surpasses that of PIXIE by roughly an order of magnitude.)  We enter these values into equations (\ref{eq:inu_mu}) and (\ref{eq:inu_y}) to obtain the curves in Figure \ref{fig:inub}.  As can be seen in the figure, below $\nu\lesssim 500$ GHz, the intergalactic dust foreground remains below the $5\sigma$ sensitivity limit of the $y$-type distortions.  At higher frequencies, the dust foreground exceeds PIXIE's sensitivity to $y$-type distortions by $\sim 1-2$ orders of magnitude.  Contamination due to intergalactic dust is a bigger problem for $\mu$-type distortions over a broader range of frequencies.  For $\nu\gtrsim 100$ GHz, the intergalactic dust foreground exceeds the PIXIE sensitivity to $\mu$-type distortions by at least 1--3 orders of magnitude.

The upper limits for \idust~derived at $z_{\rm max}=2$ and $z_{\rm max}=10$ lie almost entirely above the PIXIE sensitivity to CMB distortions at the $5\sigma$ level.   For instance, the curve representing an upper limit on IGM dust emission since $z_{\rm max}=2$ is $\sim 1-2$ orders of magnitude higher than the PIXIE sensitivity to $y$-type distortions over the entire bandwidth of the instrument.  The upper limit on intergalactic dust emission since $z_{\rm max}=10$ is as much as 3 orders of magnitude higher the PIXIE sensitivity to $y$-type distortions.  

For $60\le\nu\le 600$ GHz, the frequency range of COBE/FIRAS, the intensity of the predicted dust emission is $\sim 1$--100 times weaker than the COBE/FIRAS upper limits on the spectral distortions, which likely explains why the dust foreground has hitherto eluded detection.  Thus, for both $y$- and $\mu$-type distortions, it is not possible to entirely rule out the contribution of intergalactic dust at redshifts $0.5<z_{\rm max}\le 10$ to the cumulative foreground emission.  Getting a better sense of just how much IGM dust at all redshifts contributes to the foreground emission requires precise measurements of the intergalactic DGR at higher redshifts.  Still, it is clear that in order to take full advantage of PIXIE, PRISM, and similar future experiments' ability to measure spectral distortions, a sophisticated subtraction of the foreground emission due to intergalactic dust is necessary.

Fixsen et al. (1998) used the COBE/FIRAS data to derive the extragalactic far-infrared background (FIRB) spectrum, which we overplot in Figures \ref{fig:inua} (and \ref{fig:inub}) as a red curve.  By integrating under the solid black curve in Figure \ref{fig:inua} and comparing it to the integral of the FIRB spectrum, we determine that dust in the IGM contributes a \emph{minimum} of $\sim 0.001\%$ to the FIRB in the frequency range $\nu=30-1100$ GHz.  The upper limit on \idust~at $z=2$ represents dust emission during an epoch of interest, since the cosmic star formation rate is believed to have peaked around this time (Madau \& Dickinson 2014).  Dust expelled from galaxies into the IGM during and since this active period in the Universe's history contributes \emph{no more than} $\sim 0.03\%$ to the total FIRB.  Finally, we integrate under the curve for $z=10$, representing dust production since the epoch of reionization.   We find that, at most, the contribution to the FIRB from intergalactic dust since the epoch of reionization amounts to $\sim 0.23\%$.  

\section{Conclusion and Summary}
There is a treasure trove of information to be harvested from the CMB spectrum.  Constraining CMB spectral distortions provides a unique avenue for exploring the nature of the sources in the young Universe that induced them.  Possible sources of CMB distortions include primordial black holes (e.g., Ricotti et al. 2008; Tashiro \& Sugiyama 2008; Pani \& Loeb 2013); primordial magnetic fields (e.g., Jedamzik et al. 2000; Sethi \& Subramanian 2005; Chluba et al. 2015; Kunze \& Komatsu 2014, 2015); the decay and annihilation of relic particles, including dark matter candidates (e.g., Zeldovich et al. 1972; Chluba \& Sunyaev 2012; Chluba 2013; Hu \& Silk 1993; McDonald et al. 2001); cosmic strings (Ostriker \& Thompson 1986; Sanchez \& Gignore 1989, 1990; Tashiro et al. 2012); and Silk damping (e.g., Sunyaev \& Zeldovich 1970; Barrow \& Coles 1991; Daly 1991; Hu et al. 1994; Chluba et al. 2012).

With their high sensitivity, recently proposed experiments to measure CMB distortions---such as PIXIE and PRISM---have the potential to yield many unprecedented constraints.  But only by carefully considering all possible sources of foreground contamination will we be able to unlock the full potential of these instruments and, therefore, the wealth of information concealed in the CMB spectrum.

Whereas previous studies have considered the contamination of CMB  anisotropies from Galactic interstellar dust emission (e.g., Kogut et al. 1996; Finkbeiner et al. 1999; Kogut et al. 2011a), the purpose of this work was to estimate how emission from \emph{intergalactic} dust compares to the strength of CMB spectral distortions which the next generation of experiments aim to detect.  To meet this goal, we employed models of extragalactic background radiation capable of being absorbed by IGM dust grains, which emit the processed energy as infrared light.  We used radiative transfer modeling to determine how the temperature of IGM dust, \tdust, evolves with redshift.  Finally, we used \tdust~and estimates of DGR($z$) to predict the level of emission from IGM dust at different redshifts.  Our main results are summarized as follows:
\begin{itemize}
\item For frequencies $\nu\lesssim 500$ GHz, the dust foreground remains below the PIXIE experiment's $5\sigma$ sensitivity limit of the $y$-type distortions.  For $\nu\gtrsim 500$ GHz, the dust foreground exceeds PIXIE's sensitivity to $y$-type distortions by $\sim 1-2$ orders of magnitude.  
\item For $\nu\gtrsim 100$ GHz,  contamination due to intergalactic dust exceeds the PIXIE sensitivity to $\mu$-type distortions by 1--3 orders of magnitude.
\item The intergalactic dust emission arising from all stars and galaxies since the epoch of reionization may entirely mask the PIXIE $5\sigma$ detection of CMB distortions by a maximum of $\sim 2$ orders of magnitude and $\sim 3$ orders of magnitude for $y$- and $\mu$-type distortions, respectively.
\item The contribution to the FIRB from intergalactic dust is between 0.001\% and $\sim 0.23\%$.
\end{itemize}

Whereas the dust emission in galaxies could in principle be taken out, the intergalactic dust emission is diffuse and cannot be removed easily from sky maps of the CMB.
 
\acknowledgements
We thank Jens Chluba for his helpful comments on this paper and Rachel Somerville for generously providing the outputs to her extragalactic background light model.  This work was supported by the Harvard-MIT FFL Postdoctoral Fellowship and by NSF grant AST-1312034.

\end{document}